# Graphene platelets reinforced aluminum matrix composite with enhanced strength by hot accumulative roll bonding

Jitendar K Tiwari[a], Ajay Mandal[a], N Sathish[a,b*], Venkat A N Ch[a,b], Nikhil R Gorhe[a,b], A K Singh[a,b], S A R Hashmi[a,b], A K Srivastava[a,b]

[a] Academy of Scientific and Innovative Research (AcSIR), CSIR-AMPRI Campus, Hoshangabad road, Bhopal-462026, India.
[b] CSIR-Advanced Materials and Processes Research Institute (AMPRI), Hoshangabad road, Bhopal-462026, India.

*Abstract*

Accumulative Roll Bonding (ARB) process was used to develop few-layer graphene nano platelets reinforced aluminum matrix (GNPs/Al) composite in the form of sheets. Annealed Al sheets were ARB processed up to 6 pass along with coating of graphene between stacked sheets in first and second pass. Another set was prepared with same process parameter in the absence of graphene coating. Properties of these two set of samples were analyzed to understand the effect of graphene. Further, samples were subjected to scanning electron microscopy (SEM). The cross-sectional SEM images of processed samples were analyzed on the basis of film theory of ARB process which enables micro-level mixing of material on stacked layer interface. Hence the problem of graphene agglomeration was overcome through our process. The Raman spectra at the cross-section was taken which not only show the strong interaction of GNPs with Al matrix due to increased D and D' band but also the graphene quality enhancement on the basis of single symmetric 2D band. Samples were then subjected to universal testing machine (UTM) and Vickers microhardness tester. Results showed up to ~73% increment in yield strength and ~27% increment in hardness of GNPs/Al matrix composite. Fracture surface of tensile specimen was further examined under SEM to understand the fracture mechanics at higher passes, which elucidate elongation variation and delamination behavior of stacked sheets. Deep elongated dimples with the smoothed surface in GNPs/Al composite are the cause of reduced elongation. But the final composite was having improved strength with appreciable ductility. The improvement was further justified by increment in dislocations, calculated using Williamson-Hall plot on X-ray diffraction (XRD) data. All results signify the effectiveness of the proposed technique for the development of GNPs/Al composite.

*Corresponding Author: Dr. N. Sathish, CSIR-Advanced Materials and Processes Research Institute, Hoshangabad road, Bhopal 462026, E-mail : nsathish@ampri.res.in, Ph: +91 755 2457244

**1. Introduction-** The range of industrial applications of Aluminum Matrix Composite (AMC) is wider because of their high specific properties such as lightweight, good electrical conductivity, thermal conductivity, good elastic modulus, high strength, low coefficient of thermal expansion, good wear resistance and corrosion resistance. Severe Plastic Deformation (SPD) processes are used to produce Metal Matrix Composites (MMCs) such as Accumulative Roll Bonding (ARB) [1-7], equal channel angular processing (ECAP), high-pressure torsion (HPT) [8], etc. Among the other SPD processes, ARB is the most adoptable process because of its simplicity. It consists repeatable stacking, cutting and rolling to induce the significant dislocation density. Generally, cold roll bonding process applied for fabricating of particle reinforced composite [2] and hot roll bonding process for laminates of different metals and alloys [9]. High-temperature ARB shows better bonding strength because the material will flow comparatively higher rate than cold ARB. As a consequence of lesser processing temperature of ARB than casting, the destructive interfacial reactions, porosity and inclusions are avoided.

Graphene is getting attention in the development of multifunctional MMCs since last decade because of its unique combination of mechanical [10], electrical [11] and thermal properties[12]. The incorporation of graphene in aluminum has been studied widely. There are various works have been reported on graphene aluminum (Gr/Al) composite to influence its mechanical properties such as 62%, 54.8%, 69% and 25% improvement in ultimate tensile strength (UTS) of Al by using flake powder metallurgy [13], spark plasma sintering [14], cryomilling followed by hot extrusion [15] and wet ball milling with hot pressing [16] respectively. Thermal conductivity of graphene Al 5052 composite was increased up to 15% by friction stir processing [17]. Saboori et. al. reported the 30% improvement in hardness of hot-rolled Gr/Al composite processed by powder metallurgy [18]. Although substantial improvements were observed in properties of the composite but these processes are not scalable. Current work is to produce Gr/Al composite by ARB process, on the basis of scalability and simplicity. Very few literatures are available on using nano-particles as reinforcement through ARB process. The distribution of CNT in the aluminum was confirmed by Salimi et. al. by ARB process and it was observed that only nanotubes with diameter >30 nm and >30 walls were retained after four rolling passes at 50% reduction [19]. Graphite copper composite was fabricated through ARB process up to 30 rolling passes and it was observed that graphite is reduced to few-layer graphene after 30 cycles because of high shear forces acting on graphite, an improvement in micro-hardness was reported [20]. Graphene nanosheets (GNSs) were also used as reinforcement in Cu matrix to develop ultrafine grained and high strength GNSs/Cu composite by ARB process [21]. The basic advantages of high-temperature ARB process are (a) Micro-level mixing is possible on the basis of film theory [22] (b) Applied shear force during rolling can shear off the graphene present between the layers which may increase the quality of graphene [20] (c) Composite can be produced directly in the form of sheets. Therefore, remarkable improvements in properties of graphene reinforced MMCs can be achieve by hot ARB. Also, the use of multilayer graphene is economical in compared to high-quality graphene.

In this paper, we reported the development of uniformly dispersed graphene nanoplatelets reinforced aluminum (GNPs/Al) composite by hot ARB process and evaluate the microstructure and mechanical properties. The properties were further compared with ARB processed Al i.e. without graphene to identify the effect of graphene on Al matrix. The developed composite is expected to find applications in the automotive and aerospace industries.

## 2. Materials and method-

### 2.1 Materials and processing

Commercially available pure Al sheets of dimension 80×30×3 mm$^3$ were used for the present study which composition is listed in table 1.

**Table 1 Composition of Aluminum sheet**

| Si | Fe | Cu | Mn | Mg | Ti | Al |
|---|---|---|---|---|---|---|
| 0.228 | 0.350 | 0.0051 | 0.0168 | 0.0151 | 0.0132 | Balance |

The sheets were used for the experiment after annealing at 550$^o$C for 1 hr. Multilayer GNPs (2-10nm thickness) supplied by Cheaptubes Inc, USA was used as a reinforcement. Figs 1a and 1b show the XRD and SEM micrograph of multilayer GNPs respectively. Furthermore, Raman spectrum (fig 1c) was taken to confirm the quality of GNPs.

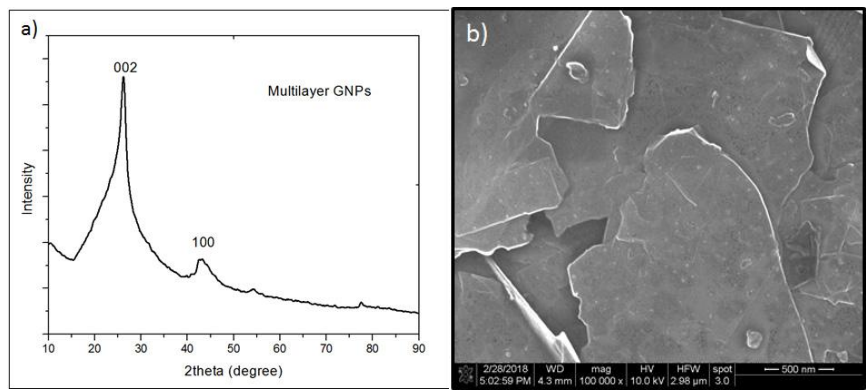

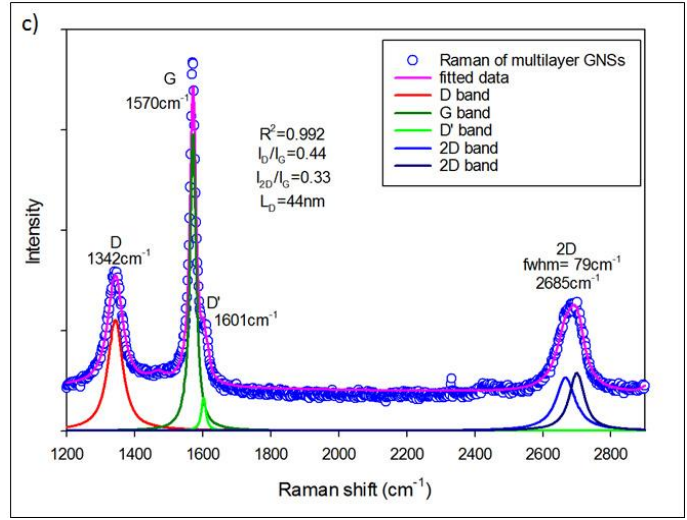

**Fig 1. a) XRD of multilayer GNPs b) FESEM image of multilayer GNPs c) Raman analysis of multilayer GNPs**

Thereafter 1g GNPs uniformly dispersed in 50 ml N-N-dimethyl formamide (DMF) solution by ultrasonication for 2hr at $50^0C$ [23]. The solution was further stirred for 3 hr. at $100^0C$ on magnetic stirrer to make slurry which was used to coat. Al sheets were wire brushed on overlapping faces to create virgin surface and then properly cleaned with ethanol. The slurry was manually coated on the same side that has been wire brushed. Finally, two sheets were stacked and clamped by Al wire to avoid misalignment during rolling. Fig 2 represents the detailed description of sample preparation. The process of degreasing, wire brushing, coating of graphene, heating and rolling was continuous i.e. there was no time lag between the subsequent processes. Two high rolling mill with power 18.65 kW was used in hot roll condition which roller diameter and roller width were 165mm and 250mm respectively. Roller temperature was maintained at $200^oC$ and roller speed was 4m/min along with 50% reduction in thickness in each pass. GNPs were added in the first and second step around 0.2 wt% in each. Third step was repeated further up to six pass for uniform dispersion. The same processing steps were followed for making ARB processed Al samples (ARBed Al). The experiment was performed in order to identify the effect of graphene on mechanical properties. The comparison of properties was made among ARB processed Al reinforced with graphene (ARBed AlwG) composite, ARBed Al and base Al sheet.

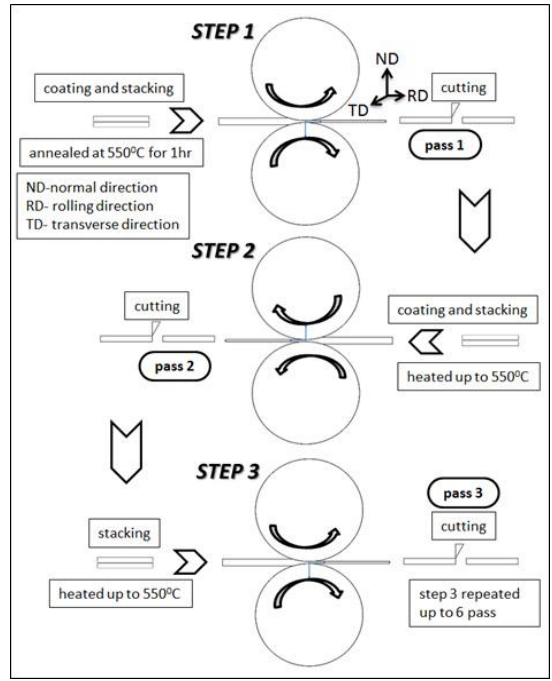

**Fig 2. Processing steps of GNPs/Al composite**

*2.2 Characterization*

The ND-RD section of ARBed AlwG was grinded and cloth polished using alumina particle suspension. Polished surface was subjected to scanning electron microscopy (SEM) in a JEOL-5600 device to observe the micro-level mixing at the interface of stacked layers. Raman spectroscopy analysis was also performed on the same surface in order to obtain the quality of graphene incorporated between the sheets after ARB process and compared with the initial Raman spectrum of multilayer GNPs. LabRam HR evolution Raman spectrometer was used for the experiment. Spectra were collected under ambient conditions using 532 nm diode laser. Further, tensile test specimen was cut from the ARBed sheet by the wire-cut electric discharge machining along the rolling direction. Size of the specimen was scaled down by 0.78 factor of subsize specimen of ASTM E8. The tensile test carried out at ambient temperature at an initial strain rate of $10^{-3}$ $s^{-1}$ using an Instron 8801 universal testing machine. The elongation in gauge length was measured manually after testing. Furthermore, X-ray diffraction was used for the calculation of dislocation density after each pass in consideration of slight change in XRD pattern due to grain refinement and strain hardening when material subjected to any SPD process. The XRD was carried out on the strained TD-RD plane by using Rigaku Miniflex II X-ray diffractometer with Cu Kα1 radiation (wavelength = 0.154056nm) in the range of $10^o$-$90^o$ with step size $0.02^o$. Williamson – Hall method was used to determine the crystallite size and dislocation density with $R^2$ value greater than 0.96. Finally, the microhardness of the polished ND-RD section of ARBed AlwG, ARBed Al and annealed Al was measured by Vickers hardness tester using 50gf for 15 sec.

## 3. Result and discussion

Table 2 represents that 6pass ARBed composite consists 64 layers of Al and 48 layers of graphene coating. To observe distribution of these layers, SEM images of 5 and 6 pass ARBed AlwG were taken at low and high magnification along ND-RD plane (fig 3) which depicts the decreasing thickness of Al stacked sheet as well as the extrusion of material between the cracks [20]. After 6 pass the average layer thickness of Al was 43 microns (46.875 microns theoretically as shown in table 2 ) which is same for ARBed Al because thickness of coated graphene in the composite is comparatively negligible.

**Table 2. Description of ARBed AlwG composite samples**

| No. of passes | Rolling temperature (°C) | No. of Al layers | No. of Graphene coating layers | Al layer Thickness (μm) |
|---|---|---|---|---|
| 1 | 550 | 2 | 1 | 1500 |
| 2 | 550 | 4 | 3 | 750 |
| 3 | 550 | 8 | 6 | 375 |
| 4 | 550 | 16 | 12 | 187.5 |
| 5 | 550 | 32 | 24 | 93.75 |
| 6 | 550 | 64 | 48 | 46.875 |

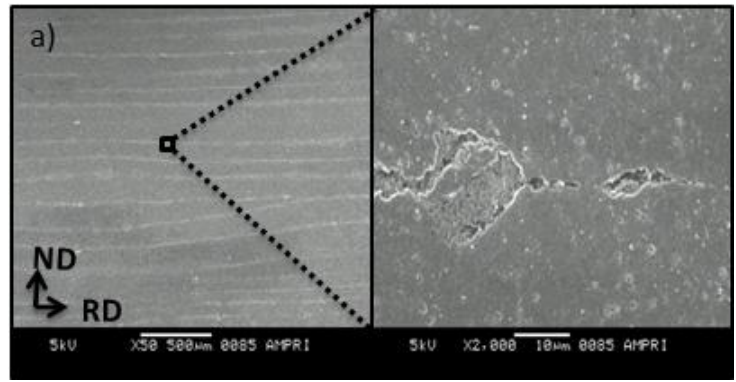

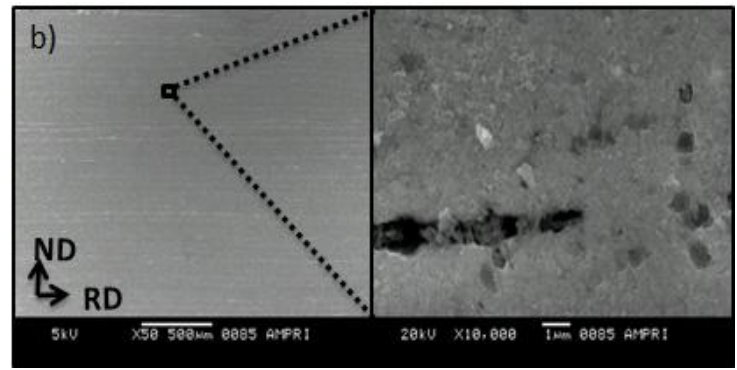

**Fig 3. a) ND-RD cross section of 5 pass AlwG  b)  ND-RD cross-section of 6 pass AlwG**

Raman spectrum shown in fig 1(c) confirms the multilayer structure of the GNPs. The $I_{2D}/I_G$ ratio is slightly higher than graphite which is the clear evidence of multilayer graphene [24]. Presence of D' band ~1600 cm$^{-1}$ indicates the existence of residual epoxy hydroxyl groups on the edges of GNPs. Raman spectra of graphene incorporated in 6pass ARBed AlwG is shown in fig 4.

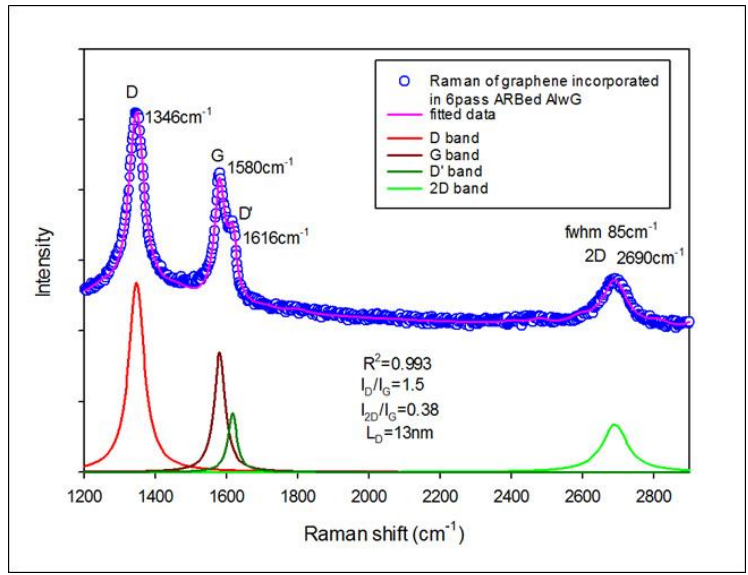

**Fig 4 Raman analysis of graphene incorporated in 6pass ARBed AlwG**

A stronger D band represents the defects in ARB processed graphene at 1346cm$^{-1}$. After rolling the D band is very well pronounced and having higher intensity than G band. Table 3 represents that defect density increased in graphene layers after ARB processing. The lateral dimension of defect-free region in graphene was calculated as

$L_D = 2.4 \times 10^{-10} (\lambda^4)(I_D/I_G)^{-1}$         [25]

Here λ is wavelength of laser i.e. 532 nm. The initial $L^D$ of multilayer GNPs was 44 nm which decreased to 12 nm in 6 pass ARBed AlwG composite. This reduction in $L^D$ indicates that the graphene surface interacted with Al matrix. G band shifted towards higher wavenumber and $I_{2D}/I_G$ value increases slightly (fig 4) after processing which indicates that graphene layers were sheared off during rolling [20, 26]. Fig 4 represents the broadened and single symmetric 2D peak after processing which was also confirmed by Ferrari et. al. that high-quality graphene gives broad, symmetric 2D peak, shifted towards higher wavenumber [27]. So it may possible that no. of graphene layers were reduced to few-layer during higher passes. The absence of characteristic peaks of graphitic carbon and graphene oxide indicate that graphene is neither agglomerated nor oxidized even though at high-temperature ARB process.

**Table 3. Calculations of distance between two point like defects ($L_D$) in initial multilayer GNSs and graphene incorporated in 6 pass ARBed AlwG composite**

| λ= 532nm | Multilayer Graphene | Graphene after 6 pass ARB |
|---|---|---|
| $I_D/I_G$ | 0.44 | 1.5 |
| $L_D$(nm) | 44 | 13 |

Tensile test was performed in order to determine the effect of graphene incorporation on tensile properties. Yield strength (YS) and ultimate tensile strength (UTS) of annealed Al, ARBed Al and ARBed AlwG are shown in figs 5a and 5b respectively. Initially, YS and UTS of annealed Al were 45.8 MPa and 105.3 MPa which increased up to 64% and 32% respectively after first pass in case of ARBed Al sample. In further passes, strength was saturated due to the saturation in strain hardening i.e. dynamic equilibrium between dislocation generation and annihilation [28]. The basic mechanisms of increased strength after the first pass were dislocation strengthening [29] and grain boundary strengthening [30]. But strength at higher passes is more sensitive to the misorientation of grain boundaries rather than the dislocation strengthening[7]. Because of heating, misorientation in grains reduces. This is probably why the substantial increment in strength was not observed even at higher passes. But intermittent heating ( up to 550 $^O$C ) was required to avoid cracking. In ARBed AlwG composite, the strengthening mechanism involved several other factors due to the incorporation of graphene nanoparticles. Around 0.4wt% graphene was added in composite after 2 pass. YS and UTS of 6 pass ARBed AlwG showed ~73% and ~37% increment respectively. This increment can be understood by following mechanisms (a) Orowan strengthening mechanism- According to this mechanism, dislocation movement restrict by the closely spaced hard nanoparticles [31]. (b) Dislocation density strengthening mechanism- in general dislocations increase in the material due to applied strain but nano-particles can also be the cause of dislocation generation because of an interfacial anomaly at the processing temperature and room temperature[6]. (c) Load transfer strengthening mechanism- nanoparticles demonstrate strong adhesion with the matrix at an interfacial region which is also an important factor to determine the mechanical behavior of the composite [32]. In case of ARBed AlwG samples, strength increased due to strain hardening at initial passes but at higher passes, strength depends on combination of above-stated strengthening mechanisms. Furthermore, elongations in tensile specimens were measured to analyze the ductile behavior. Ductility of ARBed Al initially decreased up to 3 pass and then increased with respect to number of passes (fig 5c). Elongation of the annealed Al was 60% which is two times higher than the 6pass ARBed Al. At higher passes, Al-Al bond strength increased and material density tends to theoretical level because of high applied strain i.e. 4.8 at 6 pass. In case of ARBed AlwG similar trend was observed but ductility of the composite decreased slightly may be due to interfacial bonding irregularities between graphene and Al which can act as a crack source [21]. Fig 5d depicts the increased strength along with slight reduction in ductility of the ARBed AlwG composite in comparison to 6 pass ARBed Al and annealed Al.

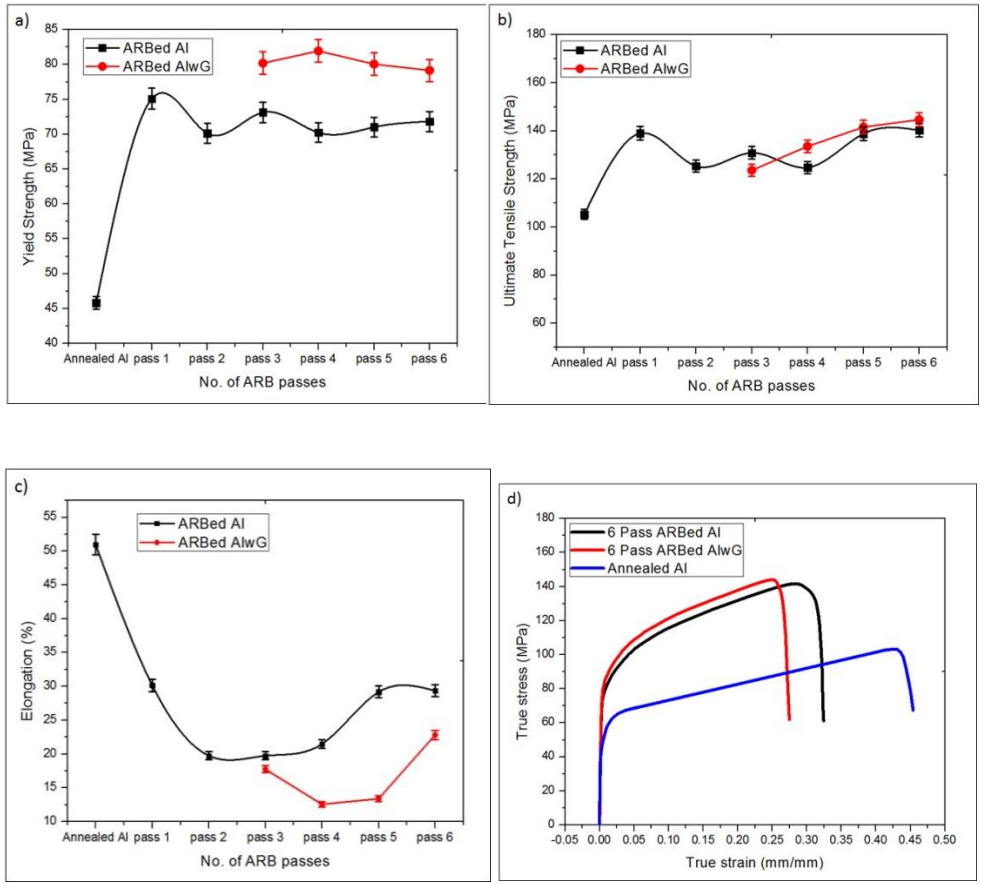

**Fig 5. a) Yield Strength vs no of ARB passes b) Ultimate Tensile Strength vs no of ARB passes c) Elongation vs no of ARB passes d) True stress vs true strain of 6 pass ARBed AlwG, 6pass ARBed Al and annealed Al**

Micro-dimples are clearly visible on the fracture surface of annealed Al which is the clear indication of ductile fracture (fig 6a). The similar hemispheroidal dimples was observed on the fracture surface of annealed Al 99.4% pure [33]. Fracture surface of 5pass and 6 pass ARBed

AlwG depicts the reduced delamination at higher passes as shown in figs 6b and 6c respectively. The concentration of microvoids was lesser at higher ARB passes but the deep and elongated dimples on fracture surface may occur due to crack initiation from weak GNPs-Al interfaces which restricted the participation of GNPs in load transfer [4].

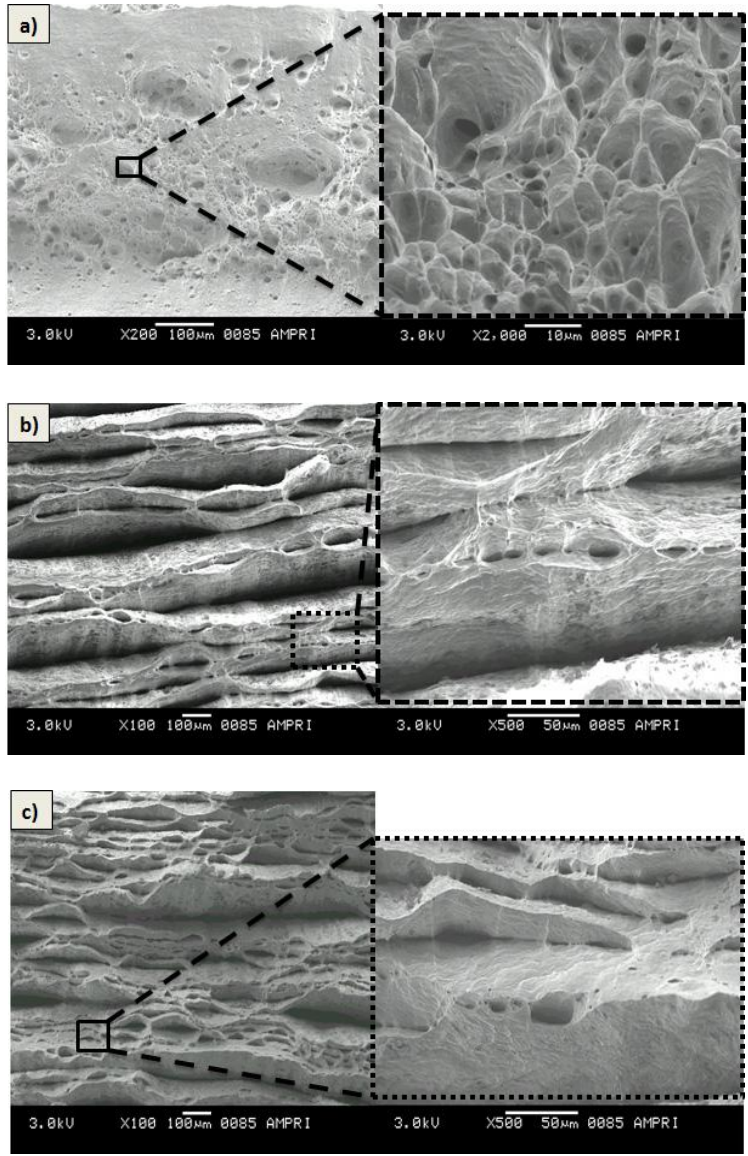

**Fig 6. a) Fracture surface of annealed Al b) Fracture surface of the 5 pass ARBed AlwG c) Fracture surface of the 6pass ARBed AlwG**

The XRD of multilayer GNPs shows the (002) diffraction line with d-spacing 0.339 nm and Bragg's angle 26.2°. Broad tail of 002 diffraction line may represent the multilayer behavior of GNPs i.e. 2-10nm thick. XRD of 1-6 pass ARBed Al samples and final AlwG composite along with annealed Al were performed on the TD-RD plane as shown in figs 7a and 7b respectively. The absence of aluminum oxide and aluminum carbide peaks confirms the impediment of destructive interfacial reactions. X-ray peak broadening was also considered for quantitative analysis of dislocations. Williamson-Hall (W-H) method was adopted to evaluate the crystallite size diameter ($l$) and lattice strain ($\varepsilon$) on the basis of full width half maxima (FWHM) measurements [5]. W-H method has been implemented by several researcher such as S.E. Shin et al[34], A. Sarkar et al[35] etc. to determine the crystallite size and lattice strain broadening when the material deform severely. In this technique, Br comprise the crystallite size broadening and lattice strain broadening which is given as-

$$Br \cos\theta = k\lambda/l + \varepsilon \sin\theta \qquad (1)$$

Where $\theta$ is bragg's angle, K is constant generally taken as unity which may consist ±10% variation and $\lambda$ is the wavelength of Cu Kα radiation i.e. 0.154056 nm. Equation 1 was used to measure the crystallite size and lattice strain by linear regression analysis (fig 7c). The mean crystallite sizes of 6pass ARBed Al and 6pass ARBed AlwG were 160nm and 110nm respectively. Calculated crystallite size is verifiable with the work on Al/$B_4C$ composite ARBed at 350°C by Moteza et. al [28]. Further dislocation density ($\rho$) is given as-

$$\rho = 2\sqrt{3}\,\varepsilon/lb \qquad (2)\ [34]$$

b= burger vector which is 0.286 nm for pure Aluminum. Dislocation densities of ARBed Al and ARBed AlwG composite in each pass were plotted against the ARB passes (fig 7d). Dislocations in 6 pass ARBed AlwG were higher than the 6 pass ARBed Al which may occur due to the thermal stresses generated near the interfacial region because of the variable interfacial characteristics at the processing temperature and room temperature[6].

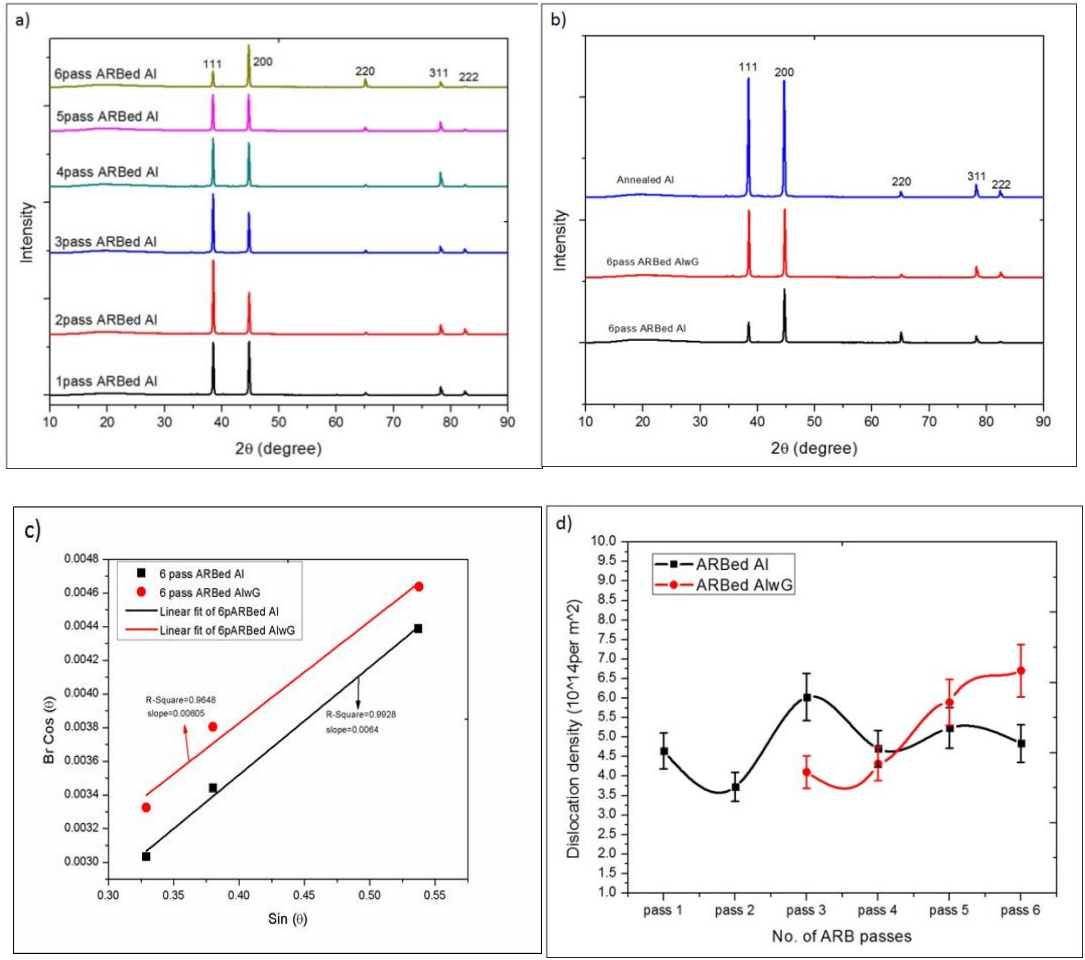

**Fig 7. a) XRD of 1-6 pass ARBed Al b) XRD of Annealed Al, 6 pass ARBed Al, 6 pass ARBed AlwG c) Williamson-Hall plot of 6 pass ARBed Al and 6 pass ARbed AlwG composite d) Dislocation density vs no of ARB passes**

Table 4 is showing the microhardness variation in ARBed Al and ARBed AlwG as a function of ARB passes. Initially, the microhardness of annealed Al was 25.4 HV which rapidly increased i.e. 33.1HV at strain 0.8 (first pass). The rapid increase in hardness at lower strain was observed due to the strain hardening which saturated at larger strain (pass 3) because of the dynamic equilibrium developed in generation and annihilation of the dislocations[28]. Strain was increasing in every pass as per 0.8n (n is the no. of ARB pass) so the stored energy within the material was increasing, therefore, material required less energy for recrystallization at high temperature rolling resulting a slight decrease in hardness of ARBed Al[36]. After fourth ARB pass, hardness of the ARBed AlwG composite was higher than the ARBed Al because at higher ARB passes the graphene was dispersed uniformly in the matrix which can participate in the enhancement of dislocations [22]. After 6pass composite showed ~27% improvement in hardness compared to base material.

**Table 4. Microhardness vs. no. of ARB passes**

| No. of ARB passes | Microhardness(HV) | |
|---|---|---|
| | ARBed Al | ARBedAlwG |
| Annealed Al | 25.4 | -- |
| Pass 1 | 33.2 | 33.1 |
| Pass 2 | 33.1 | 33.025 |
| Pass 3 | 32.42 | 31.72 |
| Pass 4 | 30.9 | 31.03 |
| Pass 5 | 30.28 | 33.53 |
| Pass 6 | 30.25 | 32.33 |

## 4. Conclusion

In this study, hot ARB process was adopted as a unique method for the development of GNPs/Al composite in the form of sheets. Microstructure, Raman, XRD studies and mechanical properties were examined and comparative study was performed among ARBed AlwG composite, ARBed Al and Annealed Al sheets. It was observed that hot ARB process can be adopted for the development of GNPs/Al composite at large scale with the synergistic effect of graphene as reinforcement. Graphene interfacial interaction in ARBed AlwG composite was confirmed by Raman analysis. Yield strength of the composite was ~73% higher than the base material along with acceptable ductility which is favorable for the structural applications. Dislocation density measurement with the help of XRD study exhibited good agreement with the experimentally observed mechanical properties of GNPs/Al composite. In conclusion, the accumulative roll bonding is a promising method for preparation of GNPs/Al composite as the reduction in no. of graphene layers and strong interface bonding can be achieved due to the process itself. A higher wt% of graphene can be incorporated successfully by increasing no. of coatings and no. of passes which can be studied further.

**Acknowledgement**

The authors sincerely thank Director, CSIR-AMPRI, Bhopal for providing experimental facilities and giving permission for publishing this work. The corresponding author wants to thank CSIR, India and AcSIR-AMPRI, Bhopal for providing the fellowship and support respectively.

**Conflict of Interest**

The authors declare that they have no conflict of interest.